\documentclass[%
 reprint,
 amsmath,amssymb,
 aps,
]{revtex4-2}

\usepackage{graphicx}
\usepackage{dcolumn}
\usepackage{bm}
\usepackage{multirow}
\usepackage{siunitx}
\usepackage{csquotes}

\graphicspath{.,{Figures/}}          
\DeclareGraphicsExtensions{.png,pdf} 

\begin{document}

\title{Turbulent Dynamics in Active Solids}

\author{Wilhelm Sunde Lie, Ingve Simonsen}
\author{Paul Gunnar Dommersnes}%
 \email{contact author: paul.dommersnes@ntnu.no}
\affiliation{%
Department of Physics, NTNU -- Norwegian University of Science and Technology, Trondheim, Norway
}%

\date{\today}

\begin{abstract}
    Turbulence is most commonly associated with high Reynolds number flow, however the framework of turbulent dynamics has been conceptually extended to many other fields,  such as magnetohydrodynamic turbulence, elastic wave turbulence in solids, and more recently to low Reynolds number active turbulence in biological fluids. Here we report a form of solid turbulent dynamics in a self-propelled two-dimensional elastic sheet. We show numerically that the polar ordering dynamics in the active elastic solid model (AES) exhibit hallmark features of turbulent dynamics: power-law scaling of the energy spectrum and non-Gaussian statistics of velocity increments. However, there is no energy cascade, in line with previous findings for active turbulence in fluids. These results extend the concept of active turbulence to solid-state active matter, and can be important for understanding collective dynamics in  biological active solids such as bacterial colonies and epithelial cell layers.
\end{abstract}

\maketitle


Active fluids constitute a broad field within non-equilibrium physics, and include topics such as for example phase transitions, active sound, liquid crystals, topological defects, and active turbulence ~\cite{marchetti2013,wensink2012}.  Active turbulence has been experimentally observed in systems such as swimming bacterial colonies or motile epithelial cell layers, which display vortices of swirling motion and power law distributed energy spectrum~\cite{wensink2012,blanch2018}. The theory of active turbulent fluid dynamics has been developed and extensively studied using both continuum flow equations and active particle simulations~\cite{wensink2012,blanch2018,amin2017,alert2022}. 

It is not the first time that the framework of turbulent dynamics has been conceptually extended beyond Navier-Stokes flow, early examples include for example magnetohydrodynamic turbulence, or acoustic and spin wave turbulence \cite{Zeldovich,LandauElectric,Zakharov}.  Elastic turbulence dynamics has also been observed in visco-elastic polymer systems~\cite{steinberg2000} and as elastic wave turbulence in solid plates~\cite{During2006}.

 Active matter systems can undergo fluid-solid phase transitions~\cite{pagonabarraga2018,henkes2011,cates2015,baskaran2013}, and there is  growing evidence that such fluid-solid transitions of cell tissues play a central role in developmental biology~\cite{campas2018,popivic2021}.  Epithelial cell layers have also been observed to act as active solid sheets that can pulsate internally or self-propel on substrate~\cite{doxzen2013,armon2021,julicher2024,lang2024,shen2025}. Active-solid theories have also recently been applied to adaptive and self-organizing robotic systems~\cite{coulais2025,saintyves2024}.  To our knowledge active solids has not been studied from a turbulent dynamics perspective.

Here we show that the Active Elastic Solid model AES)~\cite{ferrante2013,ferrante2013B}, arguably the simplest nonlinear model of a polar solid, exhibit hallmark features of turbulent dynamics:  power law scaling of the energy spectrum, and velocity increments with a strong non-Gaussian statistics at intermediate scales, similar to high Reynolds number turbulence in fluids. We also observe strongly localized polarity waves that move at constant velocity throughout the systems, reminiscent of driven domain wall dynamics in magnetic spin systems and magnetic domain turbulence~\cite{hardt2025,Zakharov}. There is also visual resemblance to fluid turbulence in the simulations in the sense that the velocity field is crowded with vortices, however unlike in fluids these vortices slow down as elastic stress builds up, and may spontaneously change direction. Such a visual resemblance to turbulence have also be seen in experiments on active solid epithelial cell layers ~\cite{lang2024}.  
 
 The specific polarity alignment mechanism in the AES model has also been subject of considerable recent interest. The mechanism was recently coined "self-alignment" as the polarity align in the direction of the local velocity field~\cite{baconnier2025}.  The self-alignment mechanism is conceptually different from the Vicsek swarm model, where particles align propulsion direction according to the propulsion direction of their neighbors~\cite{vicsek1995}.   The difference between these two alignment mechanisms is particularly clear in solids, whereas Vicsek coupling would give a linear diffusive polar dynamics in a solid, the self-alignment mechanism results in a nonlinear equation for the polarity dynamics. More recently it has been found that the AES model exhibits many interesting and unique features that are of relevance to cell biology. It was shown both theoretically and in the case of synthetic active matter that force alignment results in nonlinear rotational polarity oscillations in confined active solids~\cite{baconnier2022,menzel2025}. Solid bacterial films show exactly the same behavior with self-sustained rotating and oscillation modes that seems to be very well described by the AES model~\cite{xu2020}. In vitro experiments  show that epithetical cells form solid monolayers that behave as oscillating or self-propelled solid sheets on a substrate~\cite{doxzen2013,lang2024,shen2025}, the polar ordering dynamics included contractive elastic topological defects that again agree very well with simulations of the AES model~\cite{lang2024}. It is therefore possible that the results obtained on the AES model could have direct significance to cell biology, such as for example in wound healing dynamics.

\begin{figure*}[ht]
     \centering
     \vspace*{-0.5cm}
     \includegraphics[width=0.85\linewidth]
     {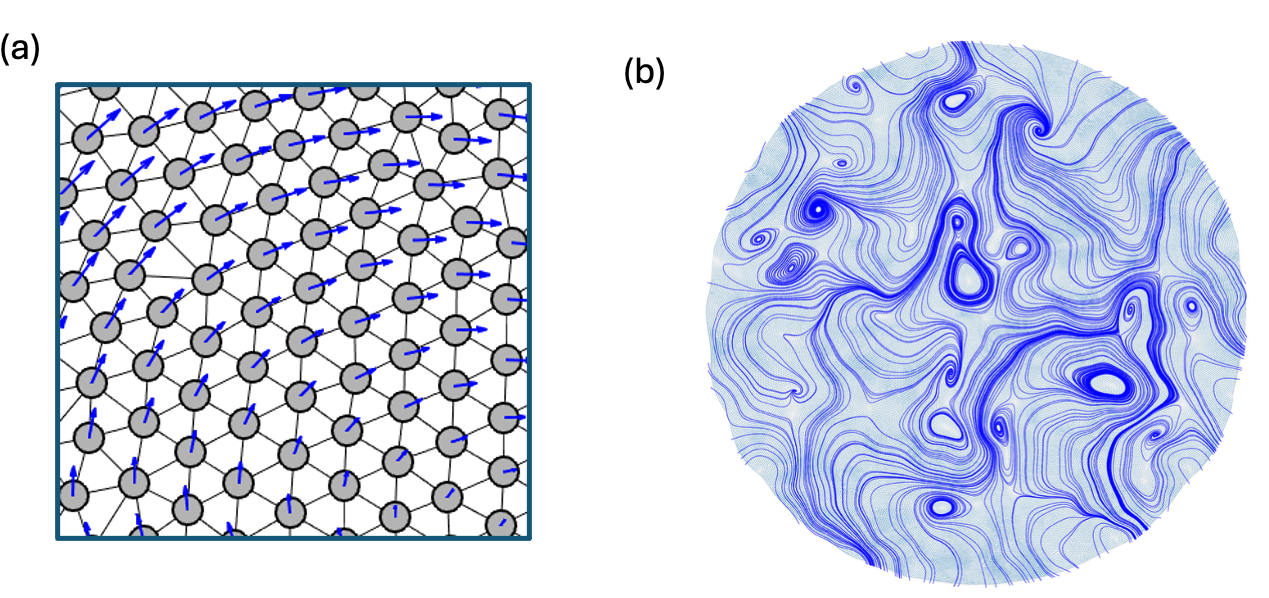}     \caption{(a)~Zoom-in showing migration velocities in the bead-spring model of an active elastic solid.  (b)~Streamlines of the velocity field of a circular patch with \num{80 000} particles. The internal active flows are possible because of the compressibility of the solid, and is dominated by vortices and their anti-vortices. Unlike   vortices in a fluid, the flow direction of vortices reverse as stress builds up. }
     \label{fig:1}
\end{figure*}


 The AES model can be expressed in continuum form (Lagrangian frame) ~\cite{ferrante2013,baconnier2022} as:
\begin{subequations}
\begin{eqnarray}
\zeta \partial_t {\bf u} &=& {\bf F}_{el}+F_a {\bf p},\label{eq:1}\\
\partial_t {\bf p}&=& -\xi {\bf p}\times({\bf p}\times {\bf F}_{el})\label{eq:2} 
\end{eqnarray}
\end{subequations}
where ${\bf u}$ is the elastic displacement field, ${\bf p}$ the polarity field, ${\bf F}_{el}$ the elastic force field. The elastic force field is given by ${\bf F}_{el}=\mu \nabla^2 {\bf u}+K\nabla(\nabla\cdot{\bf u})$ for linear elasticity. The constant $F_a$ is the magnitude of the propulsion force, and $\zeta$ is a friction coefficient, or inverse mobility coefficient. Equation~\eqref{eq:1} asserts that the speed of the particle in the absence of external forces is constant $V_a=F_a/\zeta$, and the velocity is either slowed down or increased due to external elastic forces acting on the particle. Equation~\eqref{eq:2} asserts that the polarity aligns in the direction of the total elastic force acting on the particle, with a turning rate $\xi$. This term is mathematically equivalent to the damping term in the ferromagnetic Landau-Lifshitz equation; one may thus think of the force field as equivalent to a magnetic field that orients spins in a solid.  Notice that in Eq.~\eqref{eq:2} one could also replace the elastic force ${\bf F}_i$ by $\zeta\partial_t {\bf r}_i$, \textit{i.e.}, polar alignment with elastic forces is equivalent to polar alignment with particle velocity~\cite{baconnier2025}. 

\medskip
\medskip

\noindent {\bf\large Results} \\
\noindent{ \bf Model equations.} The active elastic solid model in its simplest form~\cite{ferrante2013,baconnier2022} is given by the coupled dynamics of particle positions and particle polarity:  
\begin{subequations}
\begin{eqnarray}
   \zeta \partial_t {\bf r}_i &=&-\sum_j K(d_{ij}-r_{ij}) {\bf \hat{r}}_{ij}+F_a {\bf p}_i, 
   \label{eq:3}
   \\
    \partial_t {\bf p}_i &=&- \xi\,{\bf p}_i \times({\bf p}_i\times  {\bf F}_i),
    \label{eq:4}
\end{eqnarray}
\end{subequations}
where, for particle $i$, ${\bf r}_i$ is its position vector, and ${\bf p}_i=\left(\cos{\theta_i},\sin{\theta_i}\right)$ is its  polarity vector (propulsion direction), 
 $K$ is a linear spring constant, and $d_{ij}$ is the equilibrium distance between particle $i$ and $j$.

\begin{figure*}[ht]
    \centering
    \includegraphics[width=1.0\linewidth]{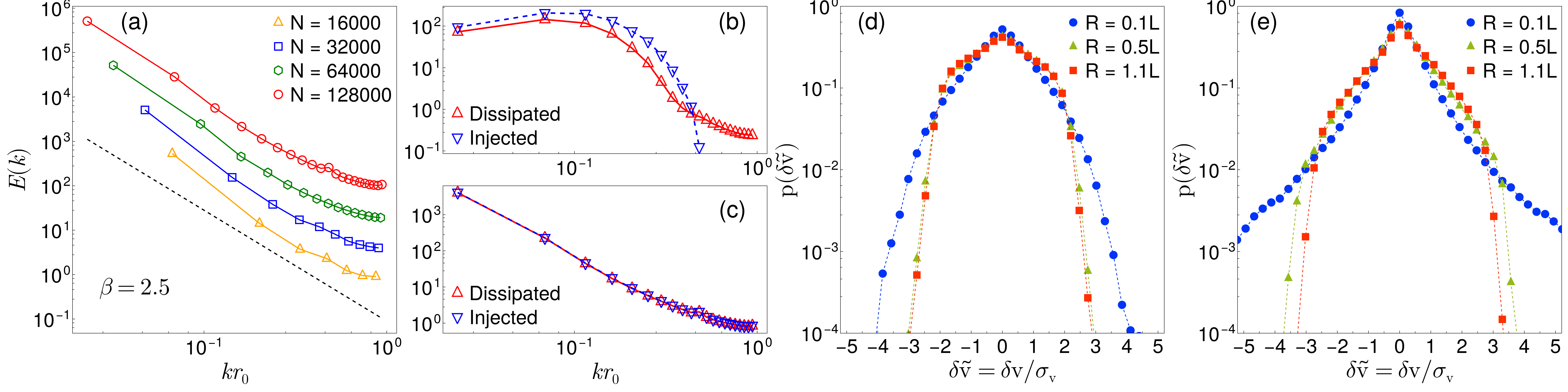}
    \caption{(a)~Energy-spectra  show broad distribution over length-scales (polar order $\Pi = 0.7$). (b)~At very early stages of ordering the injection spectra and dissipation spectra are different, indicating non-linear transfer of energy between different length scales,  however  (c)~shows that the system rapid evolves towards a state where dissipation and injection of energy are balanced on all scales, i.e. energy is injected and dissipated on the same scale. (d)~Velocity increment at polar order $\Pi = 0.1$ shows Gaussian-like behavior. (e)~As the systems evolve to $\Pi = 0.7$ there is a strong non-Gaussian velocity fluctuation on intermediate scales. The simulations were done with $N=\num{128000}$ particles,  spring constant $K=20$,  and  self-alignement turning rate $\xi=4$. The data are ensemble averages.}
    \label{fig:2}
\end{figure*}

Adding noise in the AES model can result in loss of polar order through a second-order phase transition~\cite{ferrante2013,musacchio2025}, here we choose to focus on the noise-less model, and understand the chaotic dynamics resulting form the active forces in the AES model. This is in line with  the approach to active turbulence in noiseless nematic  fluids~\cite{alert2020}.

\begin{figure*}
     \centering
     \vspace*{-0.5cm}
     \includegraphics[width=1.0\linewidth]{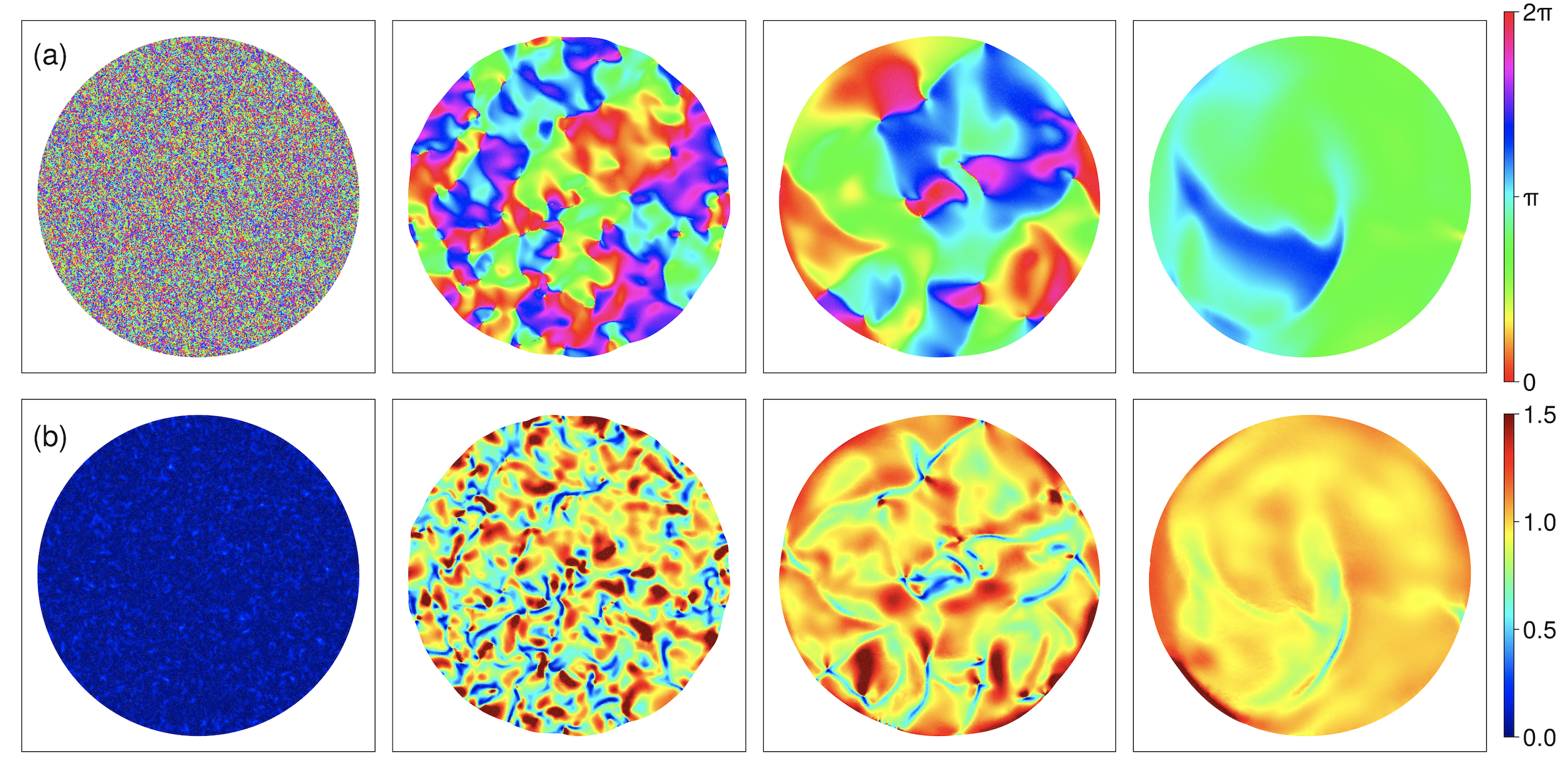}
     \caption{
     (a)~Time evolution of velocity direction (angle): the system   starts with random initial polarity directions (left) and gradually builds up into domains which move in the same direction. The initial state has polar order $\Pi = 0$ and the final state $\Pi = 0.9$. At longer times (not shown) the order becomes almost perfect  $\Pi = 1$, meaning that the patch crawls on the surface at constant speed. b)~Time evolution of the the magnitude of the velocity field (at same times as in a)). At the beginning of the simulation (left), the particles are propelled in random directions, colliding into each other, resulting in almost no motion.  Clusters  Parameters: $N=\num{128000}$, $K=20$, $\xi=6$.}
     \label{fig:1}
\end{figure*}

   The AES model is effectively a two-parameter model: rescaling length with the distance $b$ between nearest-neighbor particles ${\bf r}=\tilde{{\bf r}} b$, and time 
$t=\zeta b \tilde{t}/F_a$, results in an effective dimensionless rigidity $\tilde {K}= Kb/F_a$ and dimensionless turning rate $\tilde{\xi}=\xi K F_a/\zeta$. This amounts to setting $\zeta=1$ and $F_a=1$ in the AES equations~\eqref{eq:3} and \eqref{eq:4}. In the following, it is understood that $K$ and $\xi$ are measured in these units. We here simulate the particle AES model by direct integration of the coupled Eqs.~\eqref{eq:3} and \eqref{eq:4} starting from an initial state where the direction of the polarity vector of each particle was set randomly. In an experimental system, this corresponds to the time just after the cells are activated from a quiescent state by adding serum and start moving in initially random directions~\cite{lang2024}. The spring constant is set to $K=20$, and we considered turning rates in the range $\xi=\num{2}$--$\num{25}$. This ensures that the elastic deformations are not too large, and the linear spring model is reasonable. For $\xi < 1$, the ordering dynamics is too slow and large stress builds up, therefore, we do not consider that range here. Notice that it is straightforward to add non-linear elastic springs in this model~\cite{lang2024}, however, we choose to work with linear spring elasticity to clarify that the non-linear dynamics arises solely from the polarity-force coupling.  
  The AES model was first defined on a hexagonal bead-spring lattice~\cite{ferrante2013}, however since our study was originally motivated by understanding epithelial cell layers~\cite{lang2024} we chose to work with a statistically isotropic lattice bead-spring network. The network is created by first performing Langevin simulation of growing beads that form a disordered solid. The final bead diameters is heterogeneous and uniformly distributed in the range $b\in [0.85,1.15]$. The beads centers are then connected by Voronoi tessellation, exactly as in  references~\cite{soumya2015,lang2024}. The result is a heterogeneous network that is structurally isotropic on large scales (supplemental material). We chose to work with circular patches, with free boundary conditions. This allows for spontaneous symmetry breaking of the final polarization direction. Imposing boundary conditions is also possible, but can result in more complex dynamics due to the final state having oscillatory or unsteady dynamics.

 \noindent{ \bf Turbulent characteristics.}

\begin{figure}
    \centering
    \includegraphics[width=1.0\linewidth]{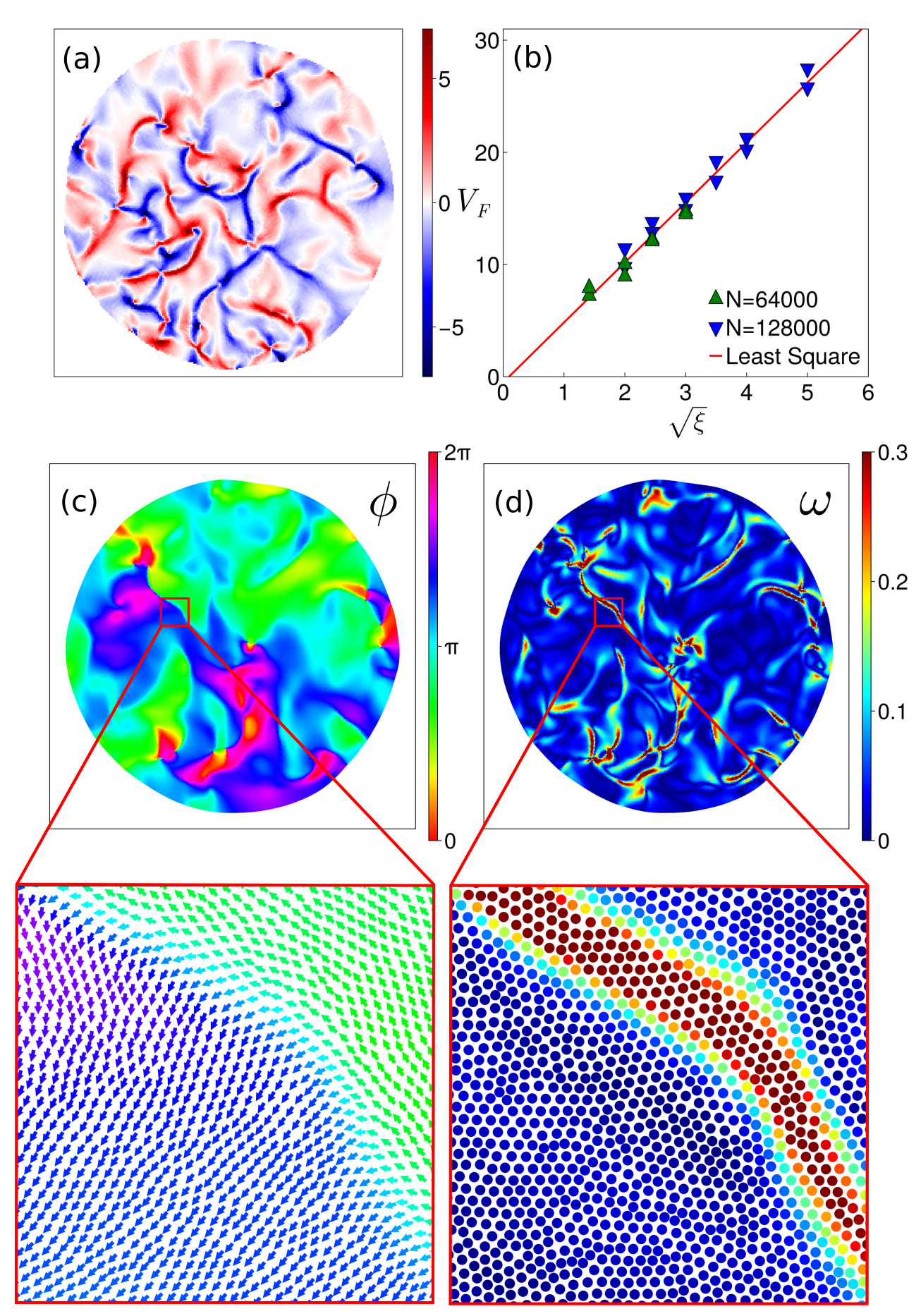}
    \caption{(a)~Velocity vorticity at $\Pi = 0.25$, with $N=\num{128000}$, $K=20$, $\xi=6$, (b)~Velocity of domain walls $V_F$ versus turning rates $\xi$. The markers are individual measurements from systems, 
    green markers originating from systems of $N = \num{64000}$ particles, while blue markers are for $N =\num{128 000}$ particles. (c)~Direction of the velocity field with cutout displaying a domain wall. (d)~Angular velocity of the velocity direction displaying high angular velocity at the domain wall.}
    \label{fig:3}
\end{figure}

The class of turbulence we shall consider is freely developing turbulence, somewhat different from steady-state turbulence usually studied in active fluids.
We are  interested in the transient dynamics of ordering, starting from an initially disordered state that evolves into global large-scale motion. It may be considered to be within the same class of phenomena as freely developing two-dimensional inertial turbulence that evolves from initially small vortices that merge into larger vortices and form a transient scale-invariant velocity field~\cite{tabeling1997}. A similar type of developing turbulence has also been found in polar active fluids~\cite{rana2020}. Our initial motivation for studying this ordering dynamics came from experiments on solid epithelial monolayers~\cite{lang2024}, where the polar ordering process takes about one day for an epithelial patch of radius \SI{1}{cm}. (This may be perceived as a slow dynamics, but as will be argued later that it should probably be considered as a very fast ordering dynamics, taking into account the slow motion of cells). In the initial stage of the ordering process the particles move in random directions, which results in a  rapid buildup of elastic forces. Small domains of polar order start to grow, which in turn drives the alignment of neighbouring domains. The polar order is measured by: $\Pi=|\sum_{i=1}^N {\bf p}_i|/N$. At the start of the simulation one has $\Pi=0$ and perfect order $\Pi \approx 1 $ is  always achieved at long times. The final ordered state is  moving in a straight line with little fluctuations. The ordering process is however not a regular coarsening process, the polar order parameter evolves in a non-monotonous and intermittent manner (supplemental Fig.~8).
 The elastic deformations of the solid are quite weak throughout the whole ordering process, as shown in Supplemental Fig.6, and would hardly evoke turbulent dynamics. However, the velocity field varies strongly both in direction and magnitude during the ordering process, as shown in Figs.~\ref{fig:1}(a) and \ref{fig:1}(b).  The velocity vorticity field shows a broad variation of structures from small to large scales, reminiscent of turbulent structures~[Fig.~\ref{fig:3}(a)]. Notice, however, that the vorticity map exhibits more string-like structures than vortices,  which means that the velocity field is dominated by fronts of sudden change of velocity, a kind of domain walls, which appear very clearly in the angular velocity field (time derivative of velocity direction), Fig.~\ref{fig:3}(d). A very similar domain wall structure was also found experimentally in solid epithelial cells during ordering~\cite{lang2024}. 
   
In active fluid turbulence one defines an \enquote{energy}~\cite{wensink2012,alert2022}: 
\begin{equation}
    E =\int \, {\rm d}^2 r \,\frac{{\bf v}^2}{2}.
     \label{eq:6} 
\end{equation}
The inertial kinetic energy does not play a direct role in these systems; this velocity square field is rather a measure of activity in a region in space. For example, some regions are jammed with frustrated orientation of cells pushing against each other, whereas others exhibit coherent motion.  The velocity energy spectrum ${\cal E}(k)$  is defined via:
$E=\int \! {\rm d} k\, {\cal E}(k)$  where $k$ is the wave number.  In a system with turbulent dynamics one expects a scale-free spectrum ${\cal E}(k) \sim 1/k^{\beta}$. We find that the AES model display such a broad distribution of energy over length-scales, with an exponent of approximately $\beta\approx 2.5$, as shown in Fig.~\ref{fig:2}(a). These energy spectra are obtained by ensemble averaging.  


In inertial hydrodynamic turbulence energy is transferred between different length-scales, this is not necessarily the case in active turbulent systems, in turbulent active nematic the active work is dissipated on the scale in which it is injected~\cite{alert2020,alert2022}. We follow Ref.~\cite{alert2020} and define an injection spectrum from the work produced by active forces: $\int {\rm d}^2 r \, F_a {\bf p}\cdot {\bf v}$, and a dissipation spectrum from  $\int {\rm d}^2 r \, \zeta\, {\bf v}\cdot {\bf v}$.  In the early state of the dynamics there is a difference between these spectra [Fig.~\ref{fig:2}(b)], showing that initially there is non-linear transfer of energy between scales, however as the system develops the two distributions become nearly identical~[Fig.~\ref{fig:2}(c)],  indicating that energy is dissipated at the scale it is injected, \textit{i.e.} there is no storage of elastic energy that is transmitted between different length-scales.

\begin{figure}
    \centering
    \includegraphics[width=1\linewidth]{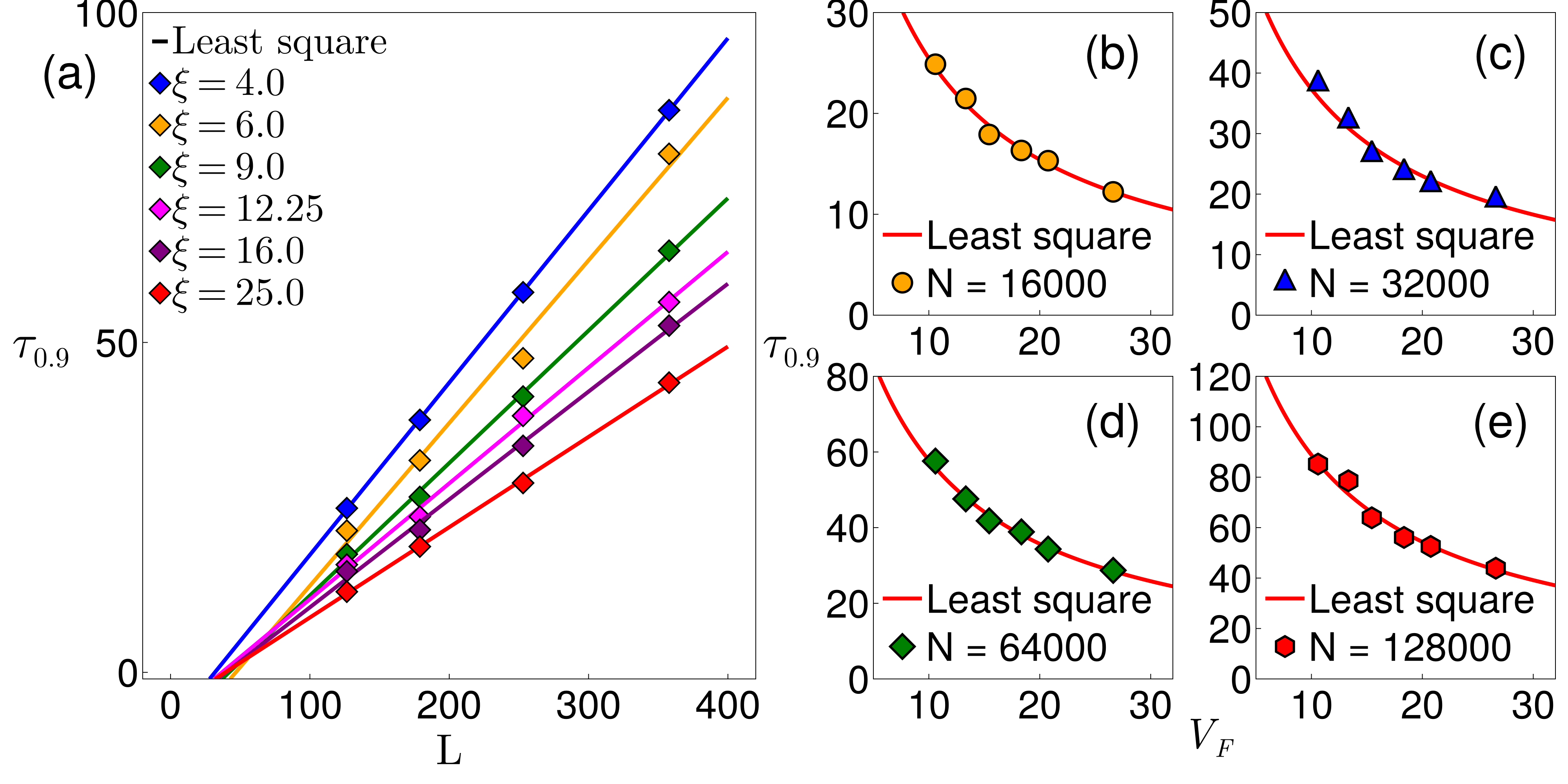}
    \caption{(a) Polarity ordering time $\tau_{0.9}$ as a function of system size $L$, for different turning rates $\xi$. Data points are accompanied by lines fitted using the least squares method. (b)--(e) Ensemble results for the time to order $\tau_{0.9}$ against the velocity of domain walls $V_F$ for different system sizes. Data points are accompanied by inverse linear, least squares fitted functions, fitted to $\tau_{0.9}^{-1}$. The polarization time is linear with system size (ballistic), and therefor much faster than the diffusive linear coarsening dynamics in a polarity-polarity coupling model.}
    \label{fig:4}
\end{figure}

A second important feature that is commonly used to characterize inertial hydrodynamic turbulence is intermittency, \textit{i.e.} sudden bursts of motion, which can be quantified by the probability distribution of the longitudinal velocity increments.
\begin{equation}
   \delta v_{\parallel}=\left[{\bf v}({\bf r}+{\bf R})-{\bf v}({\bf r})\right]\cdot {\bf \hat{R}},
   \label{eq:6} 
   \end{equation} 
where $R=|{\bf R}|$ is the distance at which velocities are compared. In inertial hydrodynamic turbulence the velocity increments are found to be non-Gaussian (broad distributions) when $R$ is an intermediate scales, and they become Gaussian on larger scales.  Such non-Gaussian velocity increments have also been observed in turbulent dynamics in granular media~\cite{radjai2002}. We find that the AES model also exhibits clear non-Gaussian velocity increments, the broadening becomes more pronounced as the system evolves towards intermediate polar order, as can be seen in Fig.~\ref{fig:2}(e).  

The AES model thus exhibits the main characteristic of turbulent dynamics, scale-free energy spectra and non-Gaussian statistics for velocity increments. 

It is not obvious that the above results would also apply to periodic boundary conditions (as is often used in active turbulence), since it would impose a global compression constraint that might reduce fluctuations, the free boundaries we use here allow global area fluctuations of the patch.

In inertial turbulence, the Reynolds number controls the transition from laminar to turbulent flow. Transition from ordered flow to turbulence is also known to occur in active fluid turbulence~\cite{amin2017,alert2022}. Although we did not explore all parameter space, we did not see any sign of a transition in the dynamics. This might have a natural explanation, there is no equivalent of viscosity in the AES model, \textit{i.e.} a term that smooths the polarity or velocity field. It is possible to combine self-alignment with Vicsek alignment, which gives diffusive (viscosity-like smoothing term)  Toner-Tu in combination with the self-alignment term~\cite{menzel2025B}, which  might result in a transition between diffusive coarsening dynamics and turbulent dynamics.

 \noindent{ \bf Domain wall dynamics}
 
 A very characteristic feature of the model is the complex dynamics of polarity domain walls. A domain wall is a sudden change in velocity (and polarity), as seen in Figs.~\ref{fig:1} and \ref{fig:3}(a).  We observed in simulations that these walls move rapid through the system at a speed that exceeds the propulsion speed of particles. If we consider that the domain wall has thickness $\ell$ and velocity $V_F$,
we can define three times:  relaxation of elastic stress over a length $\ell$: $\tau \sim \frac{\zeta \ell^2}{K}$,  characteristic turning time in stressed regions: $\tau \sim \frac{1}{\xi F_a}$, and characteristic time that particles are part of a domain wall:  $\tau\sim \frac{\ell}{V_F}$. Equating these times give scaling estimates of domain wall width:  
\begin{equation}
\ell\sim \left(\frac{K}{\zeta \xi F_a}\right)^{1/2}  
\end{equation}
and domain wall velocity:
\begin{equation}
V_F\sim \left(\frac{\xi K F_a}{\zeta} \right)^{1/2}.
\end{equation}
 where $K$ is the macroscopic elastic constant. The domain walls get sharper (less width $\ell$) and move faster with increasing
 turning rate $\xi$.

 Numerical tracking  of domain walls indeed show that their velocities on average follow a square root dependence: $V_F\sim \sqrt{\xi}$, as shown in Fig.~\ref{fig:3}(b).

 The time to produce an almost perfect polar order in the system scales linearly with system size $L$, as seen in Fig.~\ref{fig:4}(a). Notice that a similar observation was reported in Ref.~\onlinecite{ferrante2013B} for smaller systems.  This suggests a scaling relation $\tau \sim L/V_F$, \textit{i.e.} the ordering time is controlled by the speed of the domain walls.   It is noteworthy that a similar observation was recently made on domain walls dynamics in driven ferrimagnetic systems governed by the Landau-Lifshitz-Gilbert equation (where the damping term is mathematically equivalent to  the self-alignment term in the AES model), domain walls moved with a constant velocity that was proportional the square root of driving power, and there was linear relation between domain growth (coarseing) and time.~\cite{hardt2025}

  It is possible that one could consider the active solid turbulence as a kind of wave-turbulence, \textit{i.e.}, domain walls of polarity that move and collide. This behavior also highlights the difference with Vicsek-Toner-Tu type polarity coupling which would result in a linear diffusive dynamics in the solid: $\partial_t {\bf p}=\lambda \nabla^2 {\bf p}$ and hence a diffusive ordering time for reaching polar order $\tau \sim L^2$. This suggests that the AES turbulent dynamics would be a more effective means of collective ordering for cells than polarity-polarity coupling.

\smallskip
 \noindent{ \bf Conclusion}
We have shown that the transition to polar order in the AES model can be considered as a turbulent fluctuation, exhibiting power law scaling and non-Gaussian velocity increments.  The dynamics of the active solid velocity field is dominated by fronts of sudden change of velocity that propagate through the system at speeds much higher than the maximum propulsion speed of particles. Since our system is free (no boundaries) it eventually evolves into perfect order.  There appears to be no equivalent to an energy cascade in the system. In epithelial cell monolayers one would expect some non-linear elasticity, including this in the model can give an alternative mechanism of non-linear transfer of energy between scales, and possibly lead to an elastic energy cascade in the system. The results suggest that the AES model can be considered a minimal model for generating active turbulence in solids. The predictions and analysis presented here could be pursued in experiments on solid motile polar cell layers, or in synthetic active matter systems with a large number of particles.
  
\medskip
  
\noindent{ \large \bf Methods}

 The AES model was simulated using the fourth order Runge-Kutta method. The parameters for the ensemble results are detailed in Table~\ref{Tab:Parameters}. The multiple values for the turning rate $\xi$ implies a parameter sweep over these values, with all other parameters constant, meaning for systems of $N = \num{128000}$, \num{40} simulations were performed for each turning rate.
\begin{table}[h]
    \begin{center}
        \begin{tabular}{||c | c | c | c | c||} 
            \hline
            \multicolumn{5}{|c|}{Parameters} \\
            \hline
            Number of particles $N$ & \num{16000} & \num{32000} & \num{64000} & \num{128000} \\
            \hline
            \multirow{6}{*}{Turning rate $\xi$} & \num{4} & \num{4} & \num{4} & \num{4} \\
             & \num{6} & \num{6} & \num{6} & \num{6}\\
             & \num{9} & \num{9} & \num{9} & \num{9}\\
             & \num{12.25} & \num{12.25} & \num{12.25} & \num{12.25}\\
             & \num{16} & \num{16} & \num{16} & \num{16}\\
             & \num{25} & \num{25} & \num{25} & \num{25}\\
            \hline
            Spring constant $k$ & \num{20} & \num{20} & \num{20} & \num{20} \\
            \hline            
            Number of Iterations & \num{74250} & \num{104850} & \num{148500} & \num{209700} \\
            \hline
            Iterations between snapshots & \num{165} & \num{233}  & \num{330} & \num{466}  \\
            \hline
            Number of snapshots & \num{450} & \num{450} & \num{450} & \num{450} \\ 
            \hline
            Timestep $\Delta t$ & $10^{-3}$ & $10^{-3}$ & $10^{-3}$ & $10^{-3}$\\
            \hline
            Active force $F_a$ & \num{1} & \num{1} & \num{1} & \num{1} \\
            \hline
            Dissipation constant $\zeta$ & \num{1} & \num{1} & \num{1} & \num{1} \\
            \hline
            Number of simulations & \num{150} & \num{130} & \num{90} & \num{40} \\
            \hline
        \end{tabular}
    \end{center}
    \caption{A table detailing the chosen parameters for the simulations of different sizes performed for the ensemble results.}
    \label{Tab:Parameters}
\end{table}

To analyze the possible nonlinear flow of energy across different lengthscales we adopt the same approach used previously on active nematic fluid turbulence \cite{alert2020}. Multiplying the AES force balance equation and integrating over area gives:
\begin{equation}
\int \zeta {\bf v}^2 {\rm dA}=\int {\bf F}_{\rm el}\cdot {\bf v}  {\rm dA}+\int F_a {\rm p}\cdot {\bf v}  {\rm dA}.
\end{equation}
The two terms on the right of this equation represents the work done by elastic and active forces. The term on the left can be interpreted as a dissipation term.  
The dissipation and injection spectrum can be defined as:
\begin{equation}
\int \zeta {\bf v}^2 {\rm dA} = \int  {\cal D}(k) {\rm d}k
\end{equation}
\begin{equation}
 \int F_a {\rm p}\cdot {\bf v}  {\rm dA} = \int {\cal I}(k) {\rm d}k.
\end{equation}
 
Inserting  Fourier modes expansion of the velocity and polarity field:
\begin{eqnarray}
 {\bf v}({\bf r}) &=& \frac{1}{\sqrt{A}} \sum_{\bf k} {\bf V}_{\bf k} {\rm e}^{i{\bf k}\cdot {\bf r}}, \\
 {\bf p}({\bf r}) &=& \frac{1}{\sqrt{A}} \sum_{\bf k} {\bf P}_{\bf k} {\rm e}^{i{\bf k}\cdot {\bf r}},
\end{eqnarray}
into energy flow equation results in the following dissipation spectrum and energy injection spectrum:
\begin{equation}
{\cal D}(k)=\frac{1}{\Delta k}\sum_{k<|\Delta k| <k+\Delta k} \zeta  |{\bf V}_{\bf k}|^2,
\end{equation}
\begin{equation}
{\cal I}(k)=\frac{1}{\Delta k}\sum_{k<|\Delta k| <k+\Delta k}{\rm Re }\left[ F_a\, {\bf P}_{\bf k}\cdot {\bf V}_{{\bf k}}^* \right].
\end{equation}

\begin{acknowledgments}
P.G.D. thanks E. Lång, A. Lång, S. O. Bøe  and F. Brochard-Wyart for discussions on cell dynamics and active solids. We thank L. Fleinghaus, M. Nyttingnes, H. H. Haavind, T.K. Pedersen and C. Aarset Nygård  for many discussions on the AES model. We thank J.-F. Joanny for discussions and helpful suggestions on the manuscript. The simulations were performed on the Hemmer computer cluster at the Department of Physics, NTNU.  
\end{acknowledgments}


\onecolumngrid
\newpage


{\bf \large Supplemental information}

\begin{figure}[h]
    \centering
    \includegraphics[width=0.55\linewidth]{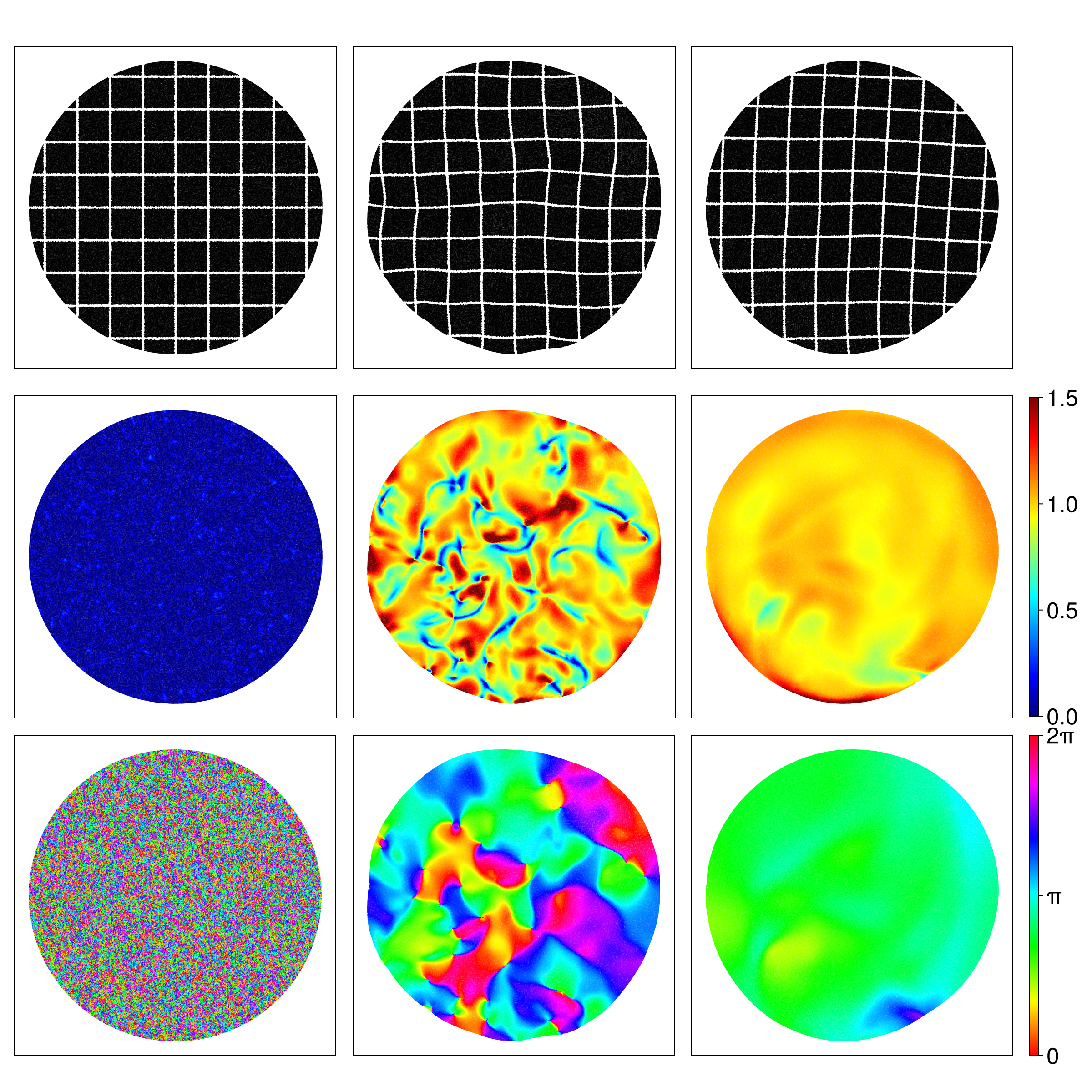}
    \caption{The extended version of Fig.~\ref{fig:1}(a)  The upper row shows the elastic deformation of the active patch.  It shows that the deformations are weak, and there is also almost no rotation of the system during the ordering process. (b) Second row shows the direction of the velocity field.  (c) Third row is the magnitude of the velocity field. Notice the large variations in velocities. The left column has an order of $\Pi = 0.0$. The center column has an order of $\Pi = 0.25$. The right column has an order of $\Pi = 0.9$. Parameters: $N=\num{128000}$, $K=20$, $\xi=6.$}
    \label{fig:ShowElasticVelPolExtended}
\end{figure}

\begin{figure}
    \centering
    \includegraphics[width=0.6\linewidth]{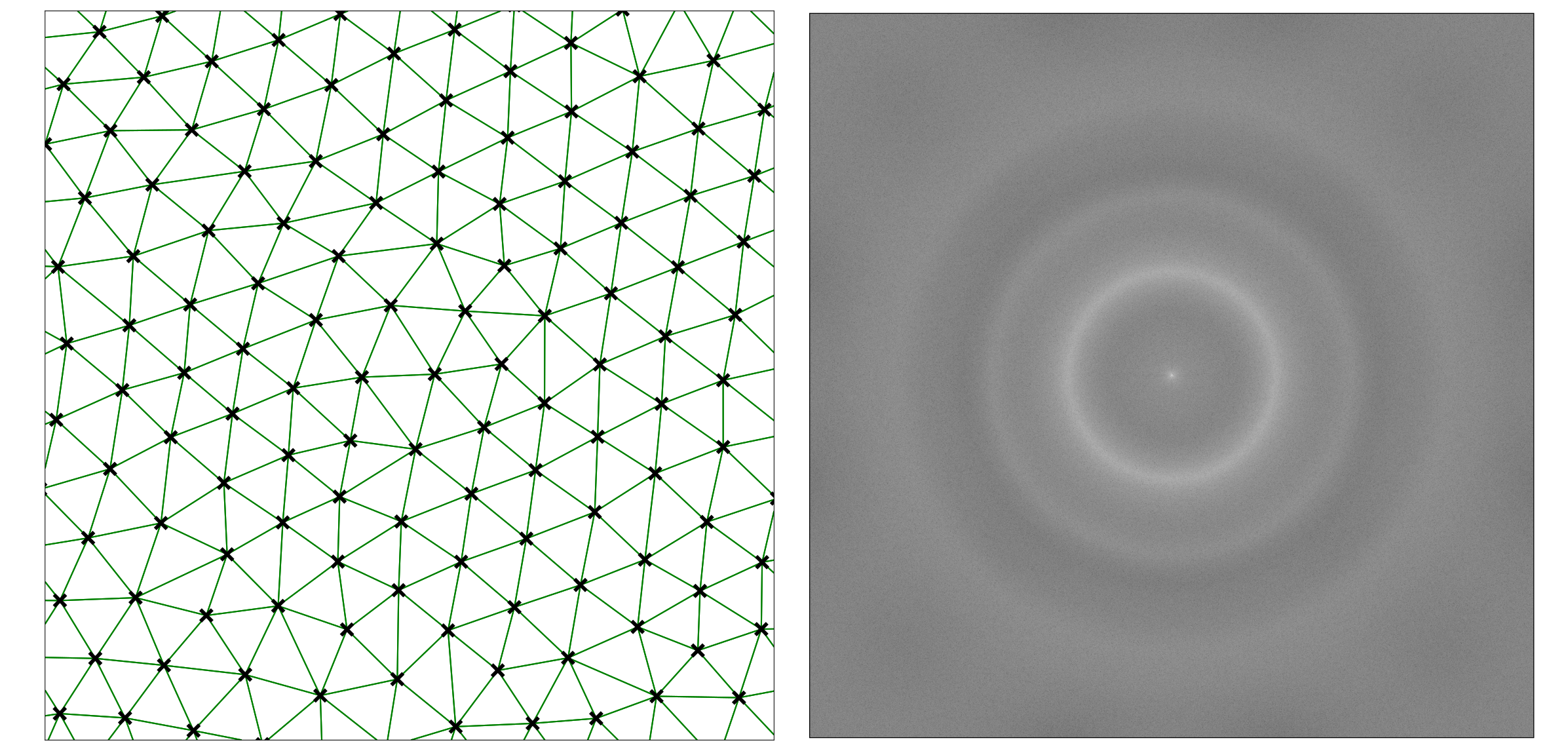}
    \caption{The left figure shows the bead-spring network derived from Delaunay neighbour pairing. The right figure shows an isotropy test of the solid: a heatmap where the values describe the logarithm of the absolute value of the 2D Fourier transform of an image of particles. }
    \label{Fig:NeighbourAndIsotropy}
\end{figure}

\begin{figure}
    \centering
    \includegraphics[width=0.6\linewidth]{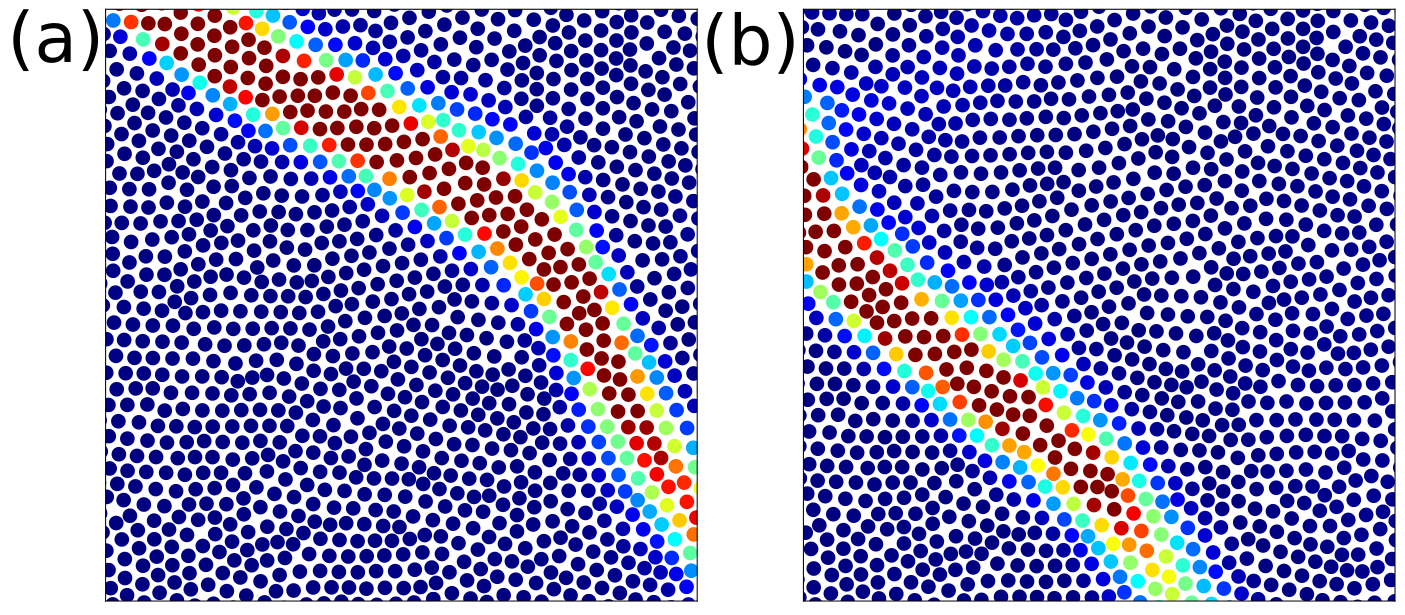}
    \caption{Polarization wave propagating through the solid. Color represents the angular velocity of polarity vector. Both figures (a) and (b) are snapshots of the same place on the solid, with only time separating the snapshots. Notice that the particles are moving very little in the process. }
    \label{Fig:MovingWave}
\end{figure}

\begin{figure}
    \centering
    \includegraphics[width=0.8\linewidth]{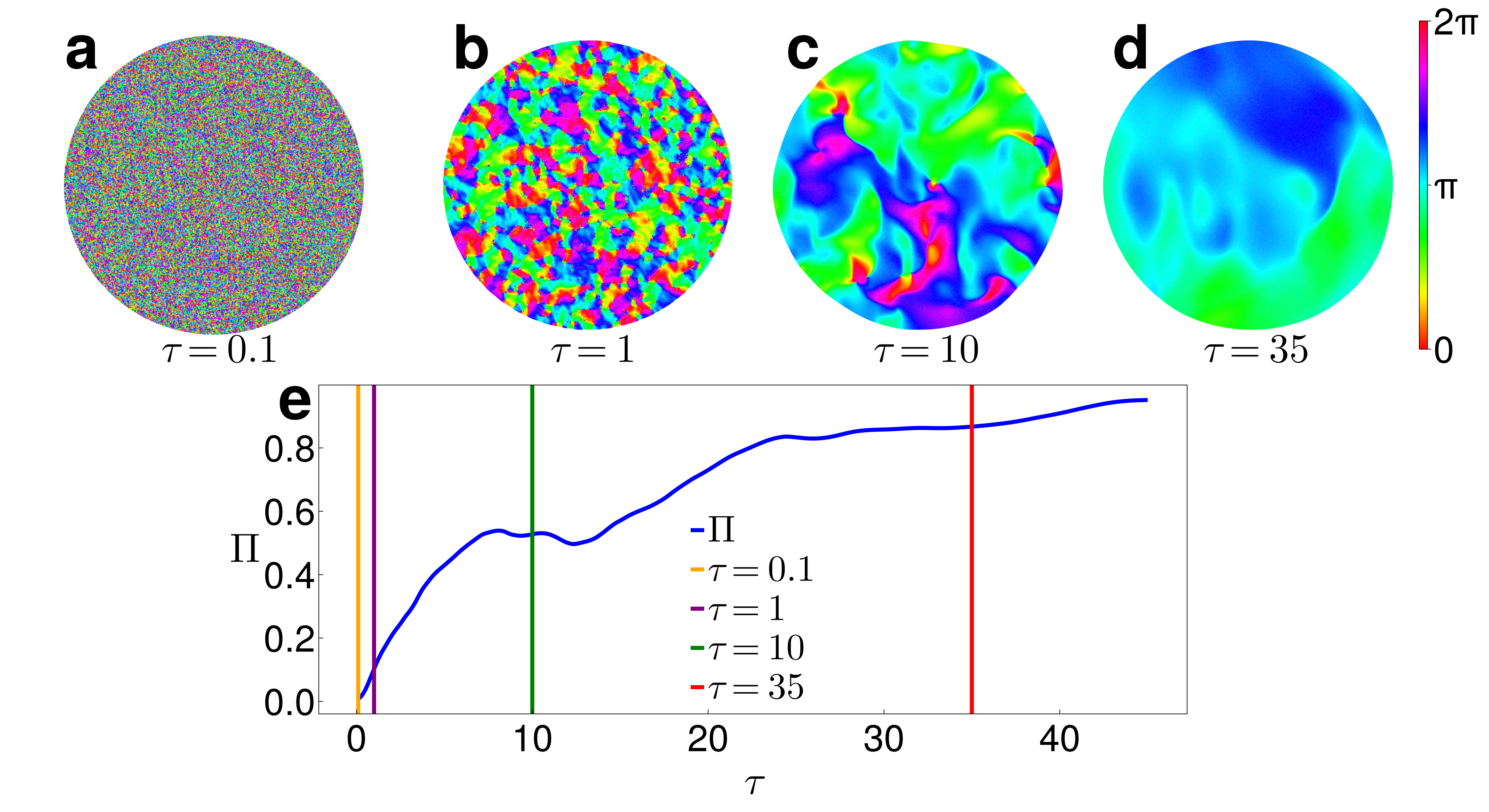}
    \caption{Polarization in the solid at four points in time \textbf{a}-\textbf{d}, along with the corresponding order parameter $\Pi$ in figure \textbf{e}. The points in figure \textbf{e} corresponding with the snapshots \textbf{a}-\textbf{d} are marked with vertical lines in \textbf{e} }
    \label{Fig:PolAngSnapBig}
\end{figure}


\end{document}